\documentclass[sigconf]{acmart}

\usepackage{booktabs} 
\usepackage{adjustbox}
\usepackage{pgfplots}
\usepackage{pdfpages}
\usepackage{soul}

\usepackage{hyperref}

\usepackage[colorinlistoftodos]{todonotes}

\acmArticle{4}
\acmPrice{15.00}
\pgfplotsset{compat=1.14}

\begin{document}

\title{Using Semi-Supervised Learning for Predicting Metamorphic Relations}

\author{Bonnie Hardin}
\affiliation{%
  \institution{Montana State University}
  \city{Bozeman}
  \state{Montana}
}
\email{bonnie.enix@montana.edu}

\author{Upulee Kanewala}
\affiliation{%
  \institution{Montana State University}
  \city{Bozeman}
  \state{Montana}
}
\email{upulee.kanewala@montana.edu}

\renewcommand{\shortauthors}{B. Hardin et al.}

\begin{abstract}
Software testing is difficult to automate, especially in programs which have no oracle, or method of determining which output is correct. Metamorphic testing is a solution this problem. Metamorphic testing uses metamorphic relations to define test cases and expected outputs. A large amount of time is needed for a domain expert to determine which metamorphic relations can be used to test a given program. Metamorphic relation prediction removes this need for such an expert. We propose a method using semi-supervised machine learning to detect which metamorphic relations are applicable to a given code base. We compare this semi-supervised model with a supervised model, and show that the addition of unlabeled data improves the classification accuracy of the MR prediction model.
\end{abstract}

%
%
%



\maketitle

\section{Introduction}
With the rapid growth of science and technology and the role it plays in the world, it is increasingly necessary to verify the accuracy of the software that produces new scientific findings. Researchers in all scientific domains face a difficult problem when it comes to testing their software. Generally, the correctness of software is determined by comparing the results of the program with the expected results. However, in the case of scientific code, the correct results are often not known; this complication, known as the oracle problem, makes testing scientific software a difficult task \cite{oracle}.

Metamorphic testing (MT) is one solution to the oracle problem. MT requires the usage of metamorphic relations (MRs) to act as an oracle for the program under test; an MR defines how a change to a test input will change the corresponding outputs. Defining MRs must often be done by the scientific domain expert, and is a time-consuming process. The greater the cost needed to built a test suite, the less likely a company or researcher is to use it. Therefore, MR prediction models are needed to decrease the amount of time and cost needed to construct an MT suite. 


There are three main categories of machine learning methods: supervised, semi-supervised, and unsupervised. In supervised machine learning algorithms, all of the data used to build the classifier have labels. These models can be unpractical because of the cost of obtaining the initial labels. In semi-supervised models, the majority of the data is unlabeled. These models becoming increasingly necessary, as big data is easily accessible on the Internet, but the corresponding labels are not. The structure of the unlabeled data helps to provide greater classification accuracy than with labeled data alone. 

In our previous work, we found that supervised learning algorithms including SVMs and decision trees are effective for predicting MRs~\cite{mrpred,mrpred2}. Our current study extends that work to include unlabeled data in the training of the binary classifiers. Unlabeled data has been shown to increase classification accuracy of machine learning models \cite{semisupervised}.
Additionally, there are many methods that do not have pre-determined metamorphic relations. These unlabeled methods can easily be added to a semi-supervised model to increase classification accuracy.

Our method uses a semi-supervised binary classification algorithm to predict metamorphic relations for methods in a program. The feature set we use consists of paths through each methods' control flow graph. These features are input to the support vector machine and label propagation algorithms, which output the predicted labels of "MR applies" or "MR does not apply". 
Our results show that the label propagation algorithm performs better than the support vector machine for 5 out of the 6 studied MRs. This result suggests the conclusion that the addition of unlabeled data, in a semi-supervised algorithm, can significantly improve on the classification accuracy of a supervised machine learning model for MR predictions.

\section{Related Work}
In the field of metamorphic relation development, several other studies have taken place recently.

A 2012 study by Liu et al. proposes a method for the composition of MRs \cite{liunew}. Their study showed that by combining two or more MRs, they can produce a new MR with a higher fault-finding effectiveness than the original. Two MRs are "compositable" if for any source test case $t_1$ and its MR $m_1$, the corresponding follow-up test case $t_2$ can be used as a source test case for a second MR, $m_2$. So $t_1$ and $t_2$ are "compositable", thereby creating a new MR.

A second paper by Su et al. studies the dynamic inference of MRs \cite{sudynamic}. The authors built a tool to implement their algorithm. The algorithm works by first defining a set of MRs, which they call transformers: "multiplier", "adder", "negator", "shuffler", and "reverser". For each transformer in the set, a function is executed with and without the transformation applied. The results are compared to see if the functions exhibit a metamorphic property. In this way, they can predict which MRs apply to a given function.

The final paper, and most similar to this study, that we mention is a 2013 study by Kanewala et al \cite{mrpred}. This study, conducted by one of the authors, has many similarities with our current work in that it uses a feature set consisting of node and path data through a function's control flow graph. These features are input into an SVM and a decision tree to build binary classifiers for metamorphic relations. The key difference between this study and ours is that this study uses supervised learning techniques, while ours extends into semi-supervised learning. Semi-supervised learning classifiers can be more accurate than their supervised counterparts because of the addition of unlabeled data \cite{semisupervised}.

\section{Method}
\begin{figure*}
\centering
\includegraphics[clip, trim=0.5cm 12cm 5cm 11cm, width=1.00\textwidth]{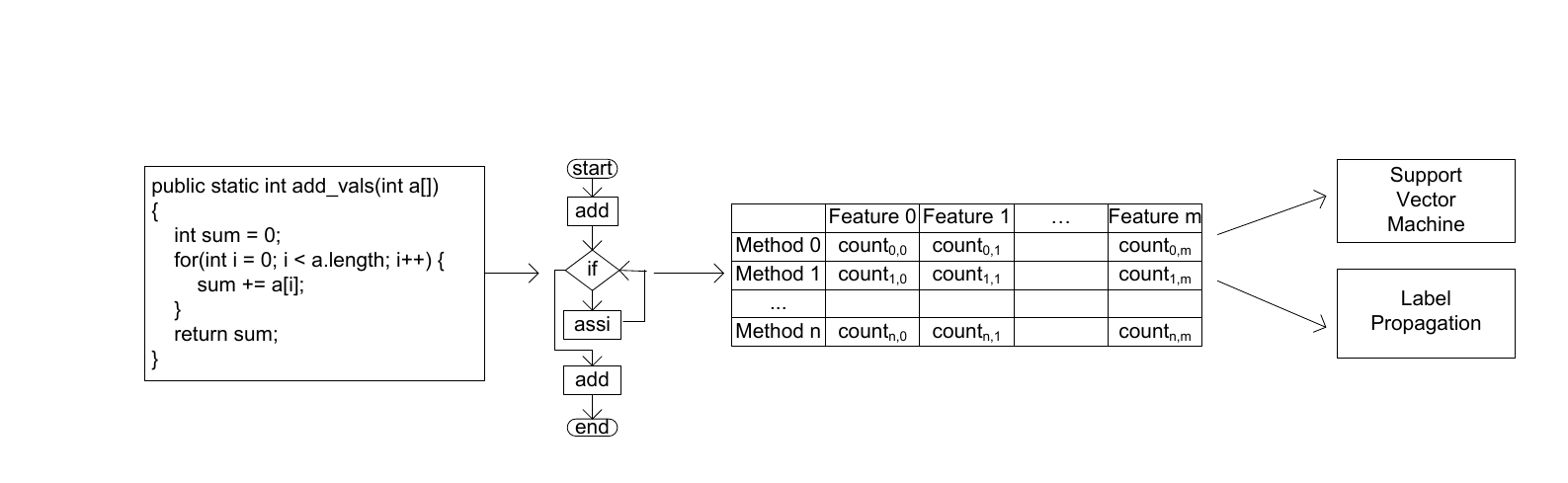}
\caption{\label{fig:experiment}Overview of Method.}
\end{figure*}

In this section, we present our method for predicting MRs using semi-supervised learning. The overview of the method is shown in Figure \ref{fig:experiment}. We begin by transforming the methods used into their control flow graphs (CFGs). We extract the features from these CFGs and input the features into the selected machine learning algorithms. These algorithms are then used to predict labels for new methods. 

\subsection{Feature Extraction}

We hypothesize there is a correlation between the paths taken through a method and the metamorphic relations found in Table \ref{tab:MRs}. To evaluate this hypothesis, we transform a set of methods into their control flow graphs, and use elements of the CFGs as features for our machine learning models. A CFG is a directed graph $G = (V, E)$, where $V$ is the set of vertices in the graph, and $E$ is the set of edges in the graph. Each vertex $v \in V$ represents an executable statement in the code. Each edge $e \in E = (v_x, v_y)$ if $x, y$ are executable statements in the code where $y$ is executed directly following $x$.

We begin with a collection of 62 open source Java methods. 
These methods all perform actions related to scientific computing, such as sorting or calculating the Hamming distance. This collection of methods was chosen as the subject of our study because of the common actions they perform in scientific computing. 

These methods were converted to control flow graphs using Soot\footnote{http://www.sable.mcgill.ca/soot/}, a Java optimization framework. These representations were stored in .dot files. After obtaining the .dot files, we extract the set of features described below from each method. The feature set consists of two types: node features and path features.   

A node feature consists of the given node, followed by the in degree and the out degree. The node features for the CFG in Figure \ref{fig:experiment} are shown in Table \ref{tab:nodefeature}. Path features consist of the shortest paths from the start node to each node in the graph, and the shortest paths from each node in the graph to the end node. The path features for the CFG in Figure \ref{fig:experiment} are shown in Table \ref{tab:pathfeature}. Paths through a control flow graph represent the possible executions of a program. Similarly, a metamorphic relation is chosen for a program based on the possible paths of execution. Therefore, we believe a feature set consisting of paths through a program is a good predictor for metamorphic relations.

\begin{table}
\centering
\begin{tabular}{|l|c|}
	\hline
     Feature & Feature Count \\
     \hline
     start - 0 - 1 & 1\\
     add - 1 - 1 & 1\\
     if - 2 - 2 & 1\\
     assi - 1 - 1 & 1\\
     add - 1 - 1& 1\\
     end - 1 - 0& 1\\
     \hline
\end{tabular}
\caption{\label{tab:nodefeature}Node features for the CFG in Figure\ref{fig:experiment}.}
\end{table}

\begin{table}
\centering
\begin{tabular}{|l|c|}
	\hline
    Feature & Feature Count\\
    \hline
    start - add & 1\\
    start - add - if& 1\\
    start - add - if - assi &1\\
    start - add - if - add& 1\\
    start - add - if - add - end&1\\
    add - if - add - end&1\\
    if - add - end&1\\
    assi - if - add - end&1\\
    add - end&1\\
    \hline
\end{tabular}
\caption{\label{tab:pathfeature}Path features for the CFG in Figure\ref{fig:experiment}.}
\end{table}

We also collect the labels for each element in our data set. For each MR under consideration, a data point is assigned a label of either 1 or 0, if the MR applies or does not apply to the method. Our experiment uses six MRs, shown in Table \ref{tab:MRs}. For each test case, $i$ is an integer value, and $c$ is a constant that is applied to $i$ to change the value for the follow-up test case. These MRs were chosen because they are commonly found in scientific computing applications. 

\begin{table}
\centering
\begin{tabular}{|l|c|r|r|}
    \hline
    MR & Initial Test Case & Follow-Up Test Case\\
    \hline
    Addition & $i_1,i_2,...,i_n$ & $i_1+c,i_2+c,...,i_n+c$\\
    Multiplication & $i_1,i_2,...,i_n$ & $i_1*c,i_2*c,...,i_n*c$\\
    Permutation & $i_1,i_2,...,i_n$ & $i_n,i_1,...,i_2$\\
    Inclusion & $i_1,i_2,...,i_n$ & $i_1,i_2,...,i_n+1$\\
    Exclusion & $i_1,i_2,...,i_n$ & $i_1,i_2,...,i_n-1$\\
    Inversion & $i_1,i_2,...,i_n$ & $1/i_1,1/i_2,...,1/i_n$\\
    
    \hline
\end{tabular}
\caption{\label{tab:MRs}Metamorphic Relations Used in the Experiment.}
\end{table}

\subsection{Support Vector Machine}
As a baseline approach, we use the support vector machine algorithm.  To do so, we use scikit-learn, a collection of Python machine learning modules \cite{scikit}. We selected scikit-learn's LinearSVC implementation of an SVM. An SVM is a supervised machine learning classification algorithm that finds a hyperplane among the data points that separates both classes of data. Data points are then classified into the positive or negative class based on their location in relation to the hyperplane.

\subsection{Label Propagation}
Label propagation is a semi-supervised machine learning classification algorithm \cite{labelprop}. We selected this algorithm to compare against the SVM, expecting the addition of unlabeled points to increase the accuracy of the model. 

The algorithm works as follows. The set of labeled data points is defined as $X_l = \{(x_1, y_1), ..., (x_l, y_l)\}$. $X_u = \{(x_l+1, y_l+1), ..., (x_l+u, y_l+u)\}$ is the set of unlabeled data points. $Y_l$ is the list of known labels, and $Y_u$ is the list of unknown labels. Given an input of $X_u$, $X_l$, label propagation outputs the values of $Y_u$.

The algorithm consists of three steps:
\begin{enumerate}
\item propagate $Y \leftarrow TY$
\item row normalize $Y$
\item clamp the labeled data
\end{enumerate}

$T$ is a probabilistic transition matrix, where $T_{ij}$ is the probability of jumping from node $j$ to node $i$. The value for any given $T_{ij}$ is calculated as $w_{ij}/\sum_{k=1}^{l+u}w_{kj}$, where $w_{ij}$ is a predetermined weight directly correlated with the Euclidean distance of the two nodes $i$ and $j$.

$Y$ is a label matrix that represents the label probability distributions of each data point. The dimensions are $(l+u) x C$, where $l$ is the length of the labeled data, $u$ is the length of the unlabeled data, and $C$ is the set of labels. $Y_i$ contains the probability that $i$ is assigned to each label $c \in C$.

The first step in the algorithm propagates the labels to previously unlabeled points. Then, $Y$ is row-normalized to ensure the label probabilities retain their meanings. The labeled data is clamped in the third step of the algorithm. This step is to ensure the original labels do not change.

Scikit-learn implements two semi-supervised algorithms: label propagation and label spreading. We selected label propagation because it clamps the original true labels; label spreading allows for the input label distributions to change over the course of the algorithm. Because our data set is relatively small, we believe that clamping the known labels will yield higher classification accuracy. 

To build the label propagation classifier, we start with the 62 data points that have calculated feature sets and labels. We randomly choose half of the data points to be considered unlabeled. 

\begin{table}[!h]
\centering
\begin{tabular}{|p{0.1\textwidth}|p{0.25\textwidth}|p{0.05\textwidth}|}
    \hline
    Parameter & Description & Value\\
    \hline
    n\_neighbors & number of neighbors used for the knn kernel & 3\\
    max\_iter & max number of iterations allowed & 1\\
    tol & threshold to consider the system at steady state & 1e-10\\
    \hline
\end{tabular}
\caption{\label{tab:parameters}Parameters used in label propagation.}
\end{table}

\section{Evaluation Method}
To evaluate SVM classifier, we split the data into training and testing sets using stratified cross validation. The training set consists of 80\% of the original data set, leaving the testing set with the remaining 20\%. The SVM builds a classifier to predict labels for previously unseen data points. An SVM takes a parameter $c$, which represents the penalty parameter of the error term; we built our model with $c=1.0$.
To to evaluate the label propagation classifier, we also use stratified cross validation, this time splitting the data into training, validation, and testing sets. To find the optimal parameters for the label propagation algorithm, we built a nested cross-validation method. The training set consists of 80\% of the total data set, and the testing set consists of the remaining 20\%. The training set then consists of  60\% unlabeled data and 40\% labeled data. The parameters accepted by the scikit-learn implementation of label propagation are shown in Table \ref{tab:parameters}.
We then test each of the models using the validation set. We select the best performing model based on accuracy score. Then, that model is tested using the test data. This process is repeated 5 times, and the scores from each run averaged together.

\section{Results and Discussion}
\begin{figure}[!h]
\begin{tikzpicture}
	\begin{axis}[
     		ybar,
            xtick=data,
            enlargelimits=0.10,
            legend style={at={(0.5,-0.15)},
            anchor=north,legend columns=-1},
            symbolic x coords={Add., Mult., Perm., Inv., Exc., Inc.},
            ylabel={Accuracy},
            nodes near coords,
            nodes near coords align={vertical},
          ]
            \addplot[ybar,fill=blue] coordinates {
                (Add., .441)
                (Mult., .717)
                (Perm., .525)
                (Inv.,  .708 )
                (Exc., .655)
                (Inc., .617)
             };
            \addplot[ybar,fill=red] coordinates {
            	(Add., .498)
                (Mult., .760)
                (Perm., .625)
                (Inv., .808)
                (Exc., .660)
                (Inc., .658)
            };
            \legend{SVM, Label Propagation}
	\end{axis}
\end{tikzpicture}
\caption{\label{fig:results}SVM and Label Propagation Results}
\end{figure}
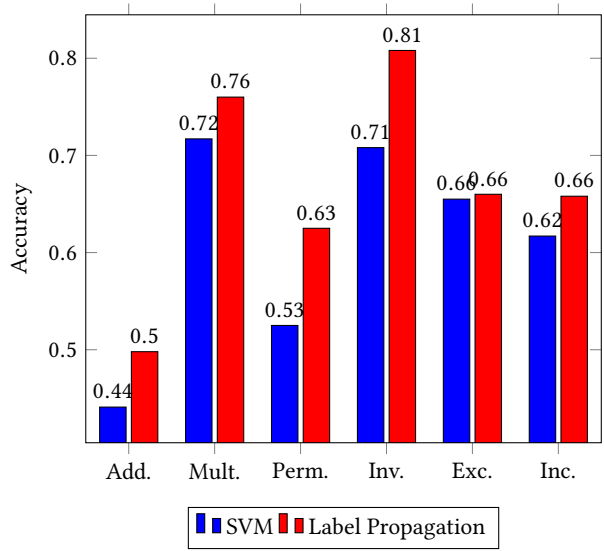

The results of the SVM and label propagation algorithms are shown in Figure \ref{fig:results}. We performed our method on the six selected MRs in Table \ref{tab:MRs}. For 5 out of the 6 MRs, the accuracy of the semi-supervised label propagation model is better than that of the supervised SVM model. 

We performed a t-test to determine the statistical relevance of the accuracy improvements. The results are shown in Table \ref{tab:ttest}. Inversion, Inclusion, and Addition have p-values of less than 0.05, representing a statistically significant change. In our previous supervised learning study, Inversion performed significantly worse than in this study. For this MR, it is clear that the addition of unlabeled data improves the prediction accuracy. 

For the MRs whose p-values represent a non-significant improvement in accuracy, we believe that the addition of more unlabeled data points would lower the p-values. We believe the small size of our data set is the key reason why these p-values are high. Improving a model with additional unlabeled data is much easier and more practical than to improve one with additional labeled data. For this reason, we believe our method to be a promising approach to use in the future for predicting MRs.

\begin{table}[!h]
\centering
\begin{tabular}{|l|c|}
    \hline
    MR & p-value\\
    \hline
    Inversion & 0.00362 \\
    Inclusion & 0.00400 \\
    Exclusive & 0.34344\\
    Addition & 0.05369 \\
    Permutation & 0.03538 \\
    Multiplication & 0.12683 \\   
    \hline
\end{tabular}
\caption{\label{tab:ttest}T-test Comparing SVM and Label Propagation.}
\end{table}

\section{Threats to Validity}
The main threat to validity for this study is in terms of external validity. The key issue is that of generalizing based on small-scale results. The results of this study suggest label propagation as an effective algorithm for predicting metamorphic relations. On relatively simple, open-source methods, this algorithm has been effective, and has improved upon the efficiency of manual metamorphic relation generation. However, these results cannot definitively prove this method will scale to industrial sized software, especially in a system interacting with multiple software artifacts.

\section{Conclusion and Future Work}
We have presented a technique to predict metamorphic relations from Java methods. We built a machine learning model using a support vector machine and the label propagation algorithm. Both algorithms used a feature set consisting of path data throughout the graph representations of the program under test. We found that label propagation performed better than the SVM for 5 out of the 6 MRs.
The results lead to the conclusion that unlabeled data increases the prediction accuracy of a binary metamorphic relation prediction classifier.

In the future, we would like to use the label spreading instead of label propagation. This algorithm allows the $\alpha$ parameter to be relaxed so that labels are not clamped. 
Additionally, we plan to use a semi-supervised support vector machine (S3VM) to build a model for our data. We will use a graph kernel with the S3VM to determine similarities among methods, rather than the built-in kernels used by label propagation and SVMs. Because of the promising results of the label propagation algorithm, we believe a semi-supervised support vector machine would yield a higher still classification accuracy than the SVM or label propagation models.

\begin{acks}
This work is supported by award number 1656877 from the National Science Foundation. Any Opinions, findings and conclusions or recommendations expressed in this material are those of the author(s) and do not necessarily reflect those of the National Science Foundation.
\end{acks}

\bibliographystyle{ACM-Reference-Format}
\bibliography{sample.bib}

\end{document}